\documentclass{article}

\usepackage{latexsym}
\usepackage{graphics}
\usepackage{epsfig}

\begin{document}
\title{New results for the dynamical critical behaviour of the two-dimensional
Ising model}
\author{Daniel Tiggemann\footnote{Insitute for Theoretical Physics, University 
of Cologne, 50923 Cologne, Germany, European Union. Email:
dt@thp.uni-koeln.de}}
\date{\today}

\maketitle

\begin{abstract}
Using the new supercomputer JUMP at the Research Center J\"ulich, we were able
to simulate large lattices (up to $L=2 \cdot 10^6$, meaning a new world record)
for long
times (up to $T=6000$ for $L=1.5 \cdot 10^5$). Using this data,
we examined the dynamical critical exponent $z$.
The old assumption of $z=2$ with logarithmic corrections seems very unlikely
according to our data, leaving the asymptotic value of $z \simeq 2.167$.
\\

\end{abstract}

\section*{Introduction}

Although the two-dimensional Ising model was solved exactly by Onsager in 1944
and much work has been done in this field over the last decades, some questions
still remain open. One is the dynamical critical behaviour.

When we take a lattice with initially all spins up, right at the Curie point
the magnetization $M$ decays with time as $M \propto t^{-\beta / \nu z}$,
where $\beta = 1/8$ is the exponent for the spontaneous magnetization and
$\nu = 1$ is the exponent for the correlation length; both are known exactly.

Two main suggestions have been made for the value of $z$ in Glauber kinetics:
one is
$z \simeq 2.167$ asymptotically, with simple power law behaviour 
(\cite{kalle}, \cite{stauffer1}, \cite{stauffer2}), the other
is $z = 2$ with logarithmic corrections to the power law behaviour. The latter
was suggested by Domany \cite{domany} and later by Swendsen
\cite{swendsen}. Lots of work has been done in order to rule out one
of these assumptions (e.~g. \cite{nightingale1}, \cite{nightingale2},
\cite{arjunwadkar}), but with no final result.

We want to test these suggestions using new numercial data obtained by using
the new supercomputer JUMP at the Research Center J\"ulich. We compare these
with older data.

\section*{Simulations}

Data was generated by initializing a lattice with all spins up and then
doing a Monte Carlo simulation of the Ising model with Glauber kinetics.
To obtain higher speed, multi spin coding and parallelization (via MPI) were
used. Speed on one $1.7$ GHz Power4+ processor was roughly 160 million
sites per second. Up to 512 processors were used in parallel for the largest
simulations.

Although the new supercomputer JUMP is rather large, there are some
restrictions on the size of lattices that can be simulated. We have chosen
$L=1.5 \cdot 10^5$ and $L=2 \cdot 10^6$ (the old world record for the 
two-dimensional
Ising model was $L=10^6$ \cite{linke}, \cite{stauffer2}), with periodic boundary
conditions. Thus finite
size effects should be negligibly small. Several independent runs were done
for $L=1.5 \cdot 10^5$ for averaging, 50 runs each for the random number
generators $x_{n+1}=13^{13} \cdot x_n \, \mathrm{mod} \, 2^{63}$ (called LCG($13^{13}$))
and 
the 64-bit implementation of Ziff's four-tap generator R(471,1586,6988,9689)
\cite{ziff4tap}.
For $L=1.5 \cdot 10^5$ the simulations were done up to 6000 timesteps
(full sweeps through the lattice). For the larger lattices, only considerably
smaller times were possible, due to restrictions in computing time.
For investigating finite-size effects, lattices with $L=5 \cdot 10^4$ were
simulated with LCG($16807$), again averaging over 50 runs.

The effective exponent $z$ can be determined from the $M(t)$ data by numercial 
differentiation: $-1/8z = d(\log M)/d(\log t)$.

\section*{Results}

\begin{figure}
\centerline{\epsfig{file=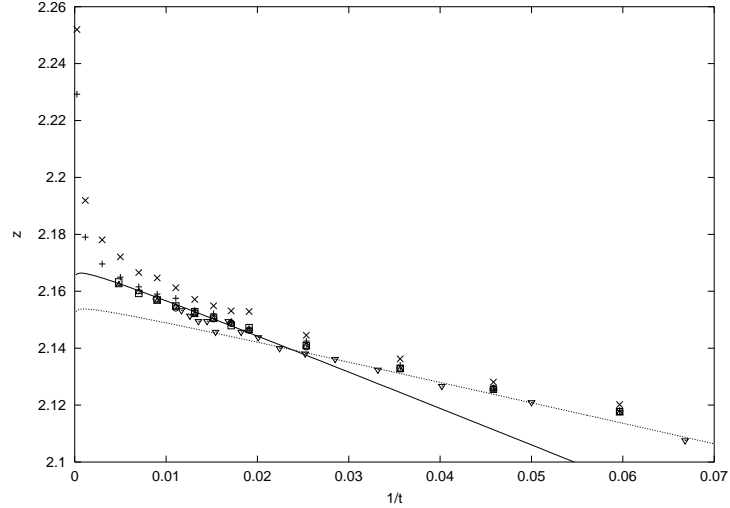, height=2.7in}}
\caption{Monte Carlo data for the two-dimensional Ising model at the
Curie point.
$+$ are for $L=1.5 \cdot 10^{5}$ with LCG($13^{13}$) generator, averaged over
50 runs,
$\times$ for $L=1.5 \cdot 10^{5}$ with R(471,1586,6988,9689) generator,
averaged over 50 runs,
$\Box$ for $L=2 \cdot 10^{6}$ with LCG($13^{13}$),
$\circ$ for $L=2 \cdot 10^{6}$ with R(471,1586,6988,9689),
$\bigtriangleup$ for  $L=2 \cdot 10^{6}$ with LCG($16807$),
$\bigtriangledown$ for $L=10^{6}$ with LCG(16807), data by
Stauffer \cite{stauffer2} (large systems one run each).
The lines represent the Swendsen suggestion for
a fit, $\beta/\left(\frac{t}{t-t_0} \cdot \left(\frac{1}{2 \beta} - \frac{c}{
1+c \cdot \log(t-t_0)}\right)\right)$, with $c=0.005$ and $t_0=0.6$ for the
solid line (new fit parameters) and $c=0.004625$ and $t_0=0.34$ for the
dotted line (Swendsen's original parameters \cite{swendsen}).}
\label{data}
\end{figure}

\begin{figure}
\centerline{\epsfig{file=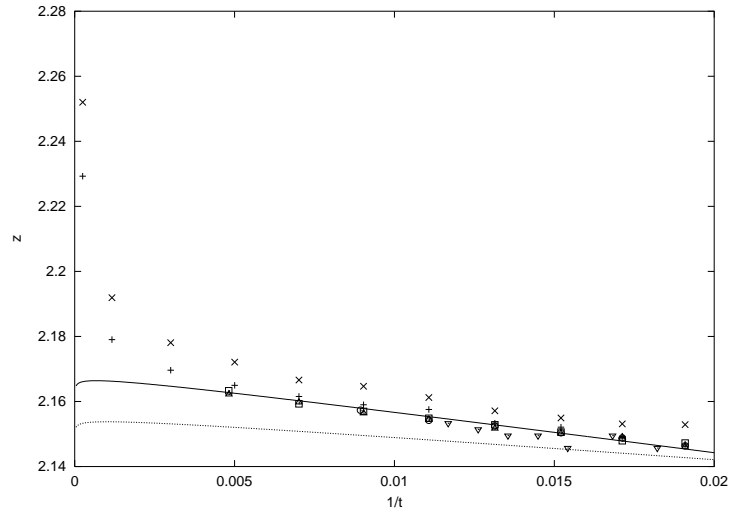, height=2.7in}}
\caption{Same plot as in fig.~\ref{data}, but with expanded $1/t$-axis.}
\label{data_snip}
\end{figure}

\begin{figure}
\centerline{\epsfig{file=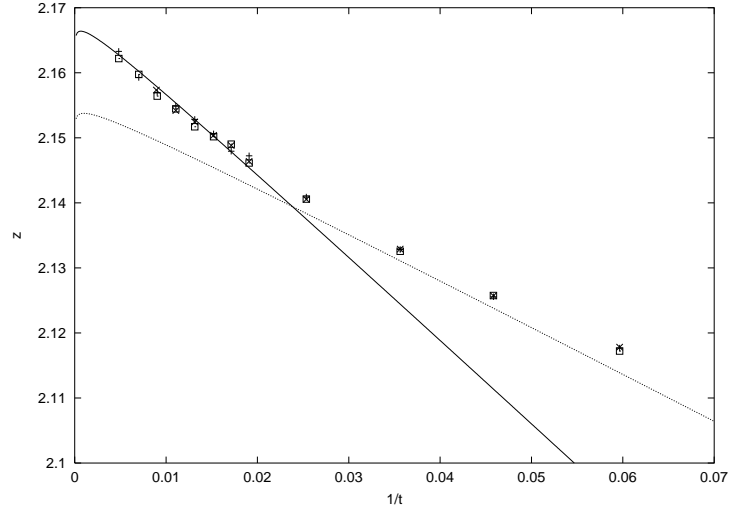, height=2.7in}}
\caption{Monte Carlo data for three runs with $L=2 \cdot 10^{6}$. $+$ are for
LCG($13^{13}$), $\times$ for R(471,1586,6988,9689), $\Box$ for LCG($16807$);
data is the same and the lines represent the same fits as in
figs.~\ref{data} and \ref{data_snip}.}
\label{l2m}
\end{figure}

\begin{figure}
\centerline{\epsfig{file=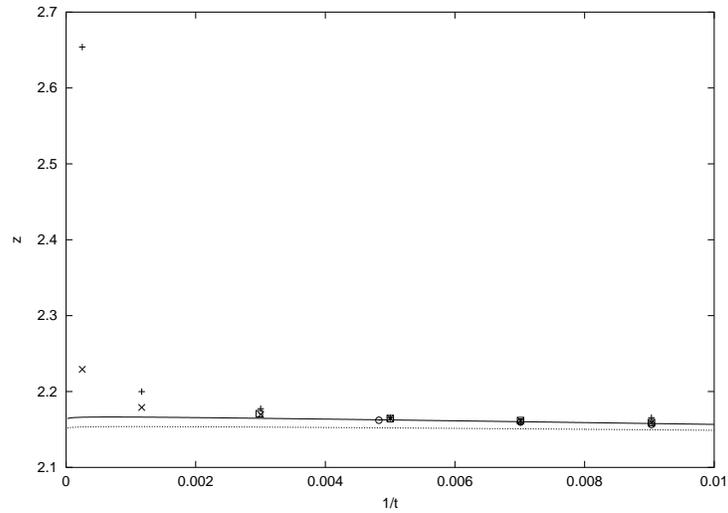, height=2.7in}}
\caption{Monte Carlo data for $L=5 \cdot 10^4$ with LCG($16807$), averaged over
50 runs ($+$),
$L=1.5 \cdot 10^5$ with LCG($13^{13}$), averaged over 50 runs ($\times$),
$L=10^6$ with R(471,1586,6988,9689), averaged over four runs ($\Box$),
and $L=2 \cdot 10^6$ with LCG($16807$), one run ($\circ$).}
\label{fin_size}
\end{figure}

Even when averaging $M(t)$ over several independent runs, fluctuations are
visible when calculating the effective $z(t)$. Thus each point in
Figs.~\ref{data}--\ref{fin_size} represents many $z(t)$; these points were
generated by dividing
the data for $z(1 \dots 6000)$ into several intervals and then doing a
least squares fit in each. Each point is the central point of the fit in the
interval. The errorbars for $z$ (not shown in the plots for better legibility)
are of
the order of the symbol size for short times and grow to up to $\pm 0.03$
for long times.
The new data, especially for the large systems, allows for a better fit of the
Swendsen suggestion. The new fitted curve has a maximum at about
$t \simeq 1700$
($1/t \simeq 6 \cdot 10^{-4}$).

The last point for $L=1.5 \cdot 10^5$, corresponding to
the interval $t=3000 \dots 6000$, is subject to strong fluctuations and thus
doubtful. Unfortunately, this is the most interesting data point. Nevertheless,
a trend is visible: for larger times, the critical exponent seems to go up,
not down, thus being in contradiction to the Domany-Swendsen suggestion.

This could also be due to finite-size effects: for $L=5 \cdot 10^4$, the effect
of increasing $z$ seems to be stronger (cf.~Fig.~\ref{fin_size}), but more
simulations would be needed for confirmation.

\section*{Conclusion}

Although it is possible to argue that the numerical data for very long times
is doubtful, as the influence of fluctuations on the value of $z$ increases,
the current precision data seems bad for the Domany-Swendsen
assumption. It would be possible to modify the fit to the
data, but the trend for long times rather contradicts the value of $z=2$.

Nevertheless, there is still work to do: the influence of various random
number generators is important, in order to find one which allows precise
data, but is still fast enough for large-scale simulations:
for $L=1.5 \cdot 10^5$, averaged over 50 runs, Ziff's four-tap generator 
produces results which differ systematically from other generators and system
sizes. This is not the case for a single run with $L=2 \cdot 10^6$ and
four-tap generator. Data for $L=5 \cdot 10^4$, averaged over 25 runs, showed
the same behavior as for $L=1.5 \cdot 10^5$. For these lattice sizes,
R(471,1586,6988,9689) seems not to be suited well.

Furthermore, the
influence of finite-size effects for long times should be investigated in
more detail, and the effects of fluctuations in the magnetization in general.

\section*{Acknowledgements}

I would like to thank D.~Stauffer for fruitful discussions and R.~H.~Swendsen
for comments on a preliminary version of this paper. I would also like to 
thank the
Research Center J\"ulich for computing time on their new supercomputer JUMP.
As a sidenote: the Ising program used for this paper was written as a simple
test for the SHMEM-library porting tool in the first place. Thus I would like
to thank David Klepacki of IBM for help with the TurboSHMEM library, although
the program now no longer uses it, but instead straight MPI.

\end{document}